\documentclass[aps,pre,twocolumn,nobibnotes]{revtex4}

\usepackage{graphicx}
\begin{document}

\title{Representative Pathways of Excitation Migration in Photosystem~I}

\author{Sanghyun Park}

\author{Melih K. \c{S}ener}

\author{Deyu Lu}

\author{Klaus Schulten}
\email[Corresponding author. E-mail: ]{kschulte@ks.uiuc.edu}

\affiliation{Beckman Institute,
University of Illinois at Urbana-Champaign, Urbana, Illinois 61801, USA}

\begin{abstract}

Photosystem I is a protein-pigment complex that performs photosynthesis
in plants, green algae, and cyanobacteria.
It contains an aggregate of chlorophylls that absorbs light and delivers
the resulting electronic excitation to the special pair of chlorophylls where
the excitation energy is used for producing charge separation 
across the cell membrane.
The seemingly random arrangement of chlorophylls in photosystem I 
poses the question
which pathways excitation migration follows towards the special pair after
absorption of light at any of its chlorophylls.
We employ a master equation to describe the process of excitation migration 
in photosystem I, and find representative paths of excitation migration 
based on the mean first-passage time from each chlorophyll to the 
special pair.
The resulting paths, beginning at each of the chlorophylls and ending at 
the special pair, provide a complete yet distilled
picture of the excitation migration towards the special pair.

\end{abstract}

\maketitle

\section*{Introduction}

Photosynthesis is carried out by pigment-protein complexes
embedded in cell membranes.
An aggregate of interacting pigments held in a fixed arrangement 
in such a complex absorbs light and delivers the
resulting electronic excitation to a reaction center, 
which uses the excitation energy 
to produce a charge separation across the cell membrane~\cite{BLAN2002}.
The transmembrane potential induced by this charge separation is later 
used for the synthesis of ATP.

Among the photosynthetic complexes, the photosynthetic unit of purple
bacteria has been most extensively studied; for a review see~\cite{Hu2002}.
Recently, a high-resolution structure of photosystem I (PSI)
has been obtained from cyanobacterium 
\textit{Synechococcus elongatus}~\cite{Jordan2001}.
PSI along with photosystem II constitutes
the machinery of oxygenic photosynthesis in plants, green algae, and 
cyanobacteria.
PSI contains an aggregate of 96 chlorophylls, 
including the special pair P700, where charge separation is initiated.
The chlorophylls are arranged without an apparent order,
except for an apparent pseudo-$C_2$ symmetry, which is in contrast
to the highly symmetric arrangement of bacteriochlorophylls
in the photosynthetic unit of purple bacteria.
The rather random arrangement of chlorophylls in PSI poses the question 
which pathways excitation migration follows towards P700
after any of its chlorophylls has absorbed light.

The rates of inter-chlorophyll excitation transfer in PSI have been 
calculated based on F\"{o}rster 
theory~\cite{BYRD2002,Sener2002,Damjanovic2002}. 
Excitation follows a stochastic trajectory along the excitation transfer
network given by these rates.
However, the obscure pattern of the excitation transfer network in PSI
(cf.\ Fig.~6 in Ref.~\cite{BYRD2002}) instills a need for 
a simpler and more distilled picture of excitation migration.
Also, the aspect that the special pair P700 is the target of the excitation 
migration process is not incorporated in the excitation transfer network
solely given by the transfer rates.
Therefore, we seek paths directed towards P700 
that are representative of all events of excitation migration 
towards P700 in a way that actual migration trajectories can be 
considered as noisy trajectories around those representative paths.

The task of finding such representative pathways is also encountered in
the so-called reaction-path problem~\cite{Elber1996,Straub2001}.
Recently, a new approach to the reaction-path problem was 
suggested~\cite{Park2002}, where mean first-passage times were
employed as reaction coordinates and reaction paths were constructed from 
them.
The suggested method is applicable to discrete systems such as 
excitation migration in photosynthetic complexes.
In this letter we apply the method to find excitation migration 
pathways in PSI\@.

\section*{Theory}

Excitation migration in PSI is a stochastic process 
governed by
the rates of inter-chlorophyll transfer, dissipation (internal conversion
of excitation to heat), and the charge separation at P700.
We build a master equation describing the process, calculate the mean first-passage time
for excitation migration from each chlorophyll to P700, 
and then find pathways representative of
the excitation migration in the excitation transfer network of the
chlorophyll aggregate towards the P700 pair.

The process of excitation migration can be described in terms of 
the probability $P_{ij}(t)$ that an excitation, initiated by light absorption
at chlorophyll $j$, resides at chlorophyll $i$ after time interval $t$.
The inter-chlorophyll transfer rates $T_{i\to j}$ 
from chlorophyll $i$ to chlorophyll $j$
are calculated as explained in Ref.~\cite{Sener2002}.
For the calculation of $T_{i\to j}$, 
we use the recently obtained site energies for 
chlorophylls~\cite{Damjanovic2002} and
the inter-chlorophyll electronic couplings given by a full 
Coulomb computation that includes all multipole contributions to the 
coupling~\cite{Sener2002}.
The dissipation rate is assumed to be the same 
($k_\mathrm{diss}\!=\!(1\,\mathrm{ns})^{-1}$) at all the chlorophylls.
Since we are here interested in paths along which excitation migrates
towards P700, 
it is sufficient to consider only first passages of excitation to P700.
(The first passage corresponds to the first term in the so-called 
sojourn expansion whose higher-order terms cover subsequent escapes and
returns to P700~\cite{Sener2002}.)
Thus, the charge-separation rate is not needed in the model, 
and the transfer rates from P700 to the other chlorophylls can be set to zero.
Collecting these rates, we obtain the master equation~\cite{Ritz2001}
\begin{equation} \label{master}
  {d\over dt}P_{ij}(t) = \sum\nolimits_{k}K_{ik}P_{kj}(t)\;,
\end{equation}
where the $96\!\times\!96$ transition matrix $K_{ij}$ is 
\begin{equation}
  K_{ij} = \bigg\{
  \begin{array}{c@{\quad\mathrm{for}\quad}l}
    T_{j\to i}-\delta_{ij}\sum\nolimits_{k}T_{j\to k}
      -\delta_{ij}k_\mathrm{diss} & j \not\in \mathrm{P700} \\
    0 & j \in \mathrm{P700}
  \end{array}\;.
\end{equation}

From the master equation (Eq.~\ref{master}) we derive a backward master 
equation which is used for calculating 
the mean first-passage time~\cite{Nadler1985,Gardiner1985}.
Since excitation migration is a Markov process, the probability $P_{ij}(t)$
can be expressed as
\begin{equation}
  P_{ij}(t) = \sum\nolimits_k P_{ik}(s)P_{kj}(t-s)
\end{equation}
for an arbitrary intermediate time $s$ ($0\!<\!s\!<\!t$).
Taking the derivative with respect to $t$ leads to
\begin{eqnarray}
  {d\over dt}P_{ij}(t) 
  &=& \sum\nolimits_k P_{ik}(s)\,{d\over dt}P_{kj}(t-s) \nonumber\\
  &=& \sum\nolimits_k P_{ik}(s) \sum\nolimits_l K_{kl}P_{lj}(t-s)\;.
\end{eqnarray}
We now take the limit of $s\to (t\!-\!0)$ and obtain the backward master 
equation
\begin{equation}
  {d\over dt}P_{ij}(t) = \sum\nolimits_{k}P_{ik}(t)K_{kj}\;,
\end{equation}
where we have used $\lim_{s\to (t-0)}P_{ij}(t-s)=\delta_{ij}$.

In the model constructed above, excitation can exit the system by two
means: the delivery to P700 and dissipation.
We emphasize again that we only consider first passages to P700.
Thus, it is convenient to consider a subsystem of 94 chlorophylls, 
excluding the P700 pair of chlorophylls.
The subsystem is described by the master equation
\begin{equation}
  {d\over dt}P_{\alpha\beta}(t) 
  = \sum\nolimits_\gamma K_{\alpha\gamma}P_{\gamma\beta}(t)
\end{equation}
or the backward master equation
\begin{equation}
\label{subbackmaster}
  {d\over dt}P_{\alpha\beta}(t) 
  = \sum\nolimits_\gamma P_{\alpha\gamma}(t)K_{\gamma\beta}\;,
\end{equation}
where Greek subscripts indicate that the P700 chlorophylls are not
included.
The $94\!\times\!94$ transition matrix $K_{\alpha\beta}$ is obtained 
from the original $96\!\times\!96$ transition matrix $K_{ij}$ 
by eliminating the rows and columns belonging to the P700 chlorophylls.

Consider an excitation initiated at chlorophyll $\alpha$ at time zero.
The probability that the excitation reaches P700 
between time $t$ and $t\!+\!dt$ is equal to 
$\sum\nolimits_\beta dt\,\xi_{\beta}P_{\beta\alpha}(t)$, where
$\xi_\beta\!=\!\sum\nolimits_{i\in\mathrm{P700}}T_{\beta\to i}$ 
is the transfer rate to P700 from chlorophyll $\beta$.
Thus, the (conditional) mean first-passage time $\tau_\alpha$, 
namely the average exit time given that the exit has occurred 
through a delivery to P700 (not through dissipation), is given as
\begin{equation}
\label{psi_alpha}
  \tau_\alpha = \int_0^\infty dt\,t
    \sum\nolimits_\beta \xi_{\beta}P_{\beta\alpha}(t) \bigg/ \phi_\alpha
\end{equation}
\begin{equation}
\label{phi_alpha}
  \phi_\alpha = \int_0^\infty dt
    \sum\nolimits_\beta \xi_{\beta}P_{\beta\alpha}(t)\;.
\end{equation}
Here $\phi_\alpha$ is the total probability that the excitation is eventually 
delivered to P700.
Since the excitation may be dissipated before it reaches P700,
$\phi_\alpha$ is less than one and the denominator in Eq.~\ref{psi_alpha} 
is necessary for normalization.
From Eqs.~\ref{subbackmaster},~\ref{psi_alpha}, and~\ref{phi_alpha},
we obtain equations for $\phi_\alpha$ and $\tau_\alpha$:
\begin{equation}
  \sum\nolimits_\alpha \phi_\alpha K_{\alpha\beta} = - \xi_\beta
\end{equation}
\begin{equation}
  \sum\nolimits_\alpha \tau_\alpha \phi_\alpha K_{\alpha\beta} 
    = -\phi_\beta \;.
\end{equation}
The solutions to these equations are given in terms of 
the inverse matrix $K^{-1}_{\alpha\beta}$ of the matrix $K_{\alpha\beta}$:
\begin{equation}
  \phi_\alpha = -\sum\nolimits_\beta \xi_\beta K^{-1}_{\beta\alpha}
\end{equation}
\begin{equation}
  \tau_\alpha 
    = -\sum\nolimits_\beta \phi_\beta K^{-1}_{\beta\alpha}\Big/\phi_\alpha\;.
\end{equation}

The mean first-passage time $\tau_i$ ($\tau_i\!=\!0$ for $i\!\in\!\mathrm{P700}$) 
is a natural measure of the progress in the 
excitation migration towards the P700 pair~\cite{Park2002};
chlorophylls associated with shorter mean first-passage times are closer 
to the target (P700) in the migration process.
As suggested in Ref.~\cite{Park2002}, we construct excitation migration paths
according to the scheme that a path going through chlorophyll $i$ chooses the
next chlorophyll $j$ such that $T_{i\to j}(\tau_i\!-\!\tau_j)$ is maximized.
The resulting paths can be considered representative because
the mean first-passage time $\tau_i$ is an average over all events of the
excitation migration from chlorophyll $i$ to P700.
For a transfer step from chlorophyll $i$ to chlorophyll $j$,
the quantity $1/T_{i\to j}$ may be interpreted as a cost, 
and $\tau_i\!-\!\tau_j$ as a gain.
The scheme then amounts to maximizing the gain-cost ratio for each step.

\section*{Results and Discussion}

\begin{figure*}  \includegraphics{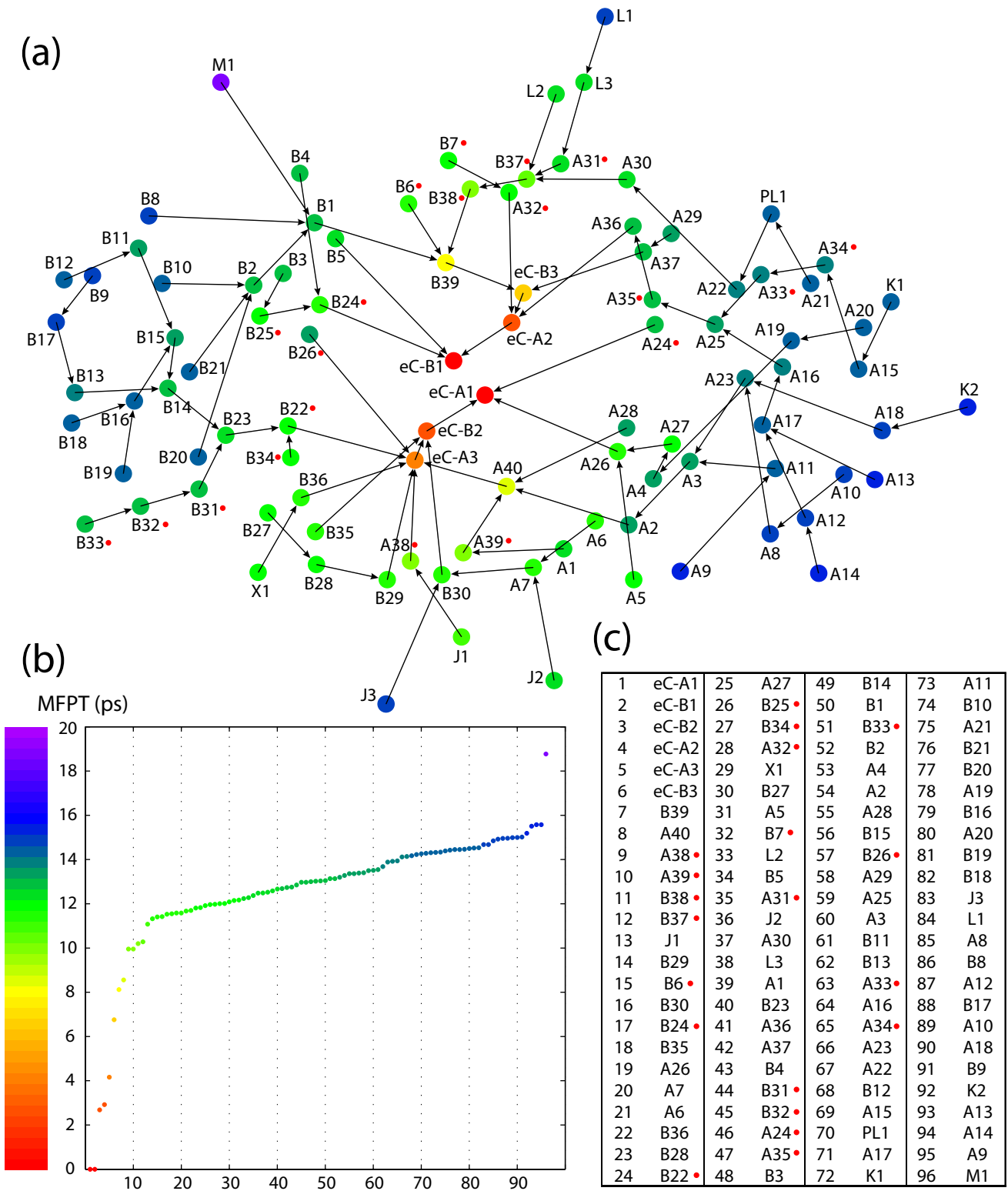}
\caption{\label{fig}
Mean first-passage time and pathways of excitation migration in PSI.
(a) The chlorophylls, projected onto the membrane plane, are denoted by
circles color-coded according to the mean first-passage time to the P700 pair.
The excitation migration paths constructed based on 
the mean first-passage time are shown as arrows.
(b) The mean first-passage time is plotted in the increasing order. The color-code scheme is
the same as in (a).
(c) The chlorophylls are sorted in the order of increasing mean first-passage
time.
The nomenclature (eC-A1, etc) follows Ref.~\cite{Jordan2001}.
`Red chlorophyll' candidates are marked with red dots in (a) and (c).}
\end{figure*}

The above method yields the mean first-passage time and the paths 
shown in Fig.~\ref{fig}.
The 96 chlorophylls are sorted in the order of increasing 
mean first-passage time and are listed in Fig.~\ref{fig}(c).
The excitation migration paths shown in Fig.~\ref{fig}(a) exhibits 
a network without an apparent order, 
as expected from the arrangement of the chlorophylls.
There are various paths, including six direct paths to P700 
(from chlorophylls eC-A2, eC-B2, A24, A26, B5, and B24) 
and the most complicated path composed of nine steps
(A21$\to$PL1$\to$A22$\to$A30$\to$B37$\to$B38$\to$B39$\to$eC-B3$\to$eC-A2$\to$eC-B1).
We here emphasize again that excitation does not
necessarily follow these paths,
but that they should be understood as representative paths.

As can be seen from Fig.~\ref{fig}(b), 
most of the chlorophylls (9th to 96th) have mean first-passage times 
around or above 10\,ps, with the average over all being 12.5\,ps.
Chlorophylls A40 and B39 lie between these peripheral chlorophylls
and the reaction-center chlorophylls which coincide with the six chlorophylls
with the shortest mean first-passage times.
This supports to a certain extent the suggestion that 
chlorophylls A40 and B39 connect the reaction-center chlorophylls and the
peripheral chlorophylls~\cite{Jordan2001}.
But, among the found excitation migration paths to P700, 
many go through neither chlorophyll A40 nor B39.
Therefore, these two chlorophylls are not the only connection between the
reaction center and the periphery, and should not be considered as 
bottlenecks.

The last chlorophyll in Fig.~\ref{fig}(c), chlorophyll M1, raises a question 
as its mean first-passage time (18.8\,ps) is much longer than the 
second longest one (15.6\,ps).
This does not imply that excitation is trapped at this chlorophyll;
18.8\,ps is still much shorter than the dissipation time of 1\,ns.
But, it seems inappropriate that one chlorophyll is located relatively 
farther from the rest.
PSI exists as a trimer in vivo~\cite{Boekema1987,Jordan2001}
and chlorophyll M1 lies close to the boundary between monomers.
By computing the electronic couplings between chlorophyll M1 and the
chlorophylls in the next monomer, we find that chlorophyll M1 is coupled to
chlorophyll A30 in the next monomer with the coupling of 52.9\,cm$^{-1}$,
which is much stronger than the strongest coupling it has 
within its own monomer (6.7\,cm$^{-1}$ with B8).
Chlorophyll M1 functionally belongs to the next monomer, not its own monomer.

In PSI, it is experimentally known that seven to eleven chlorophylls absorb 
light at longer wavelengths than the special pair P700~\cite{PALS98,ZAZU2002}.
These `red chlorophylls' have not been identified with certainty, 
but some candidates (marked with red dots in Figs.~\ref{fig}(a) and (c)) 
have been proposed~\cite{Jordan2001,BYRD2002,Damjanovic2002,Sener2002}.
Most of the candidates are found to be
located rather close to P700 in terms of 
not only spatial distance but also the mean first-passage time.
Chlorophylls A38, A39, B37, and B38 have the shortest 
mean first-passage times among 
peripheral chlorophylls (except for B39 and A40).
With regard to the representative paths shown in Fig.~\ref{fig}(a),
chlorophylls A24, A32, A38, B22, B24, and B26 are one step away from
the reaction-center chlorophylls,
and chlorophylls A39, B6, and B38 are one step away from the `connecting
chlorophylls', A40 and B39.

In conclusion, we have found representative paths of excitation migration
in PSI based on mean first-passage times.
The mean first-passage time and the paths provide
a complete yet distilled picture of the excitation migration towards P700
as illustrated in Fig.~\ref{fig}.
We expect that our methodology will be useful for various
light-harvesting complexes as more of their high-resolution 
structures become available.

\section*{Acknowledgments}

This work has been supported by National Institute of Health grant
PHS 5 P41 RR05969.

\end{document}